\documentclass[epsf]{elsart}

\usepackage[dvips]{graphicx}
\graphicspath{{./}}

\begin{document}
\begin{frontmatter}
\title{ Collective transition densities in neutron-rich nuclei}

\author[Catania]{F. Catara}, 
\author[Catania]{E.G. Lanza}, 
\author[Legnaro]{M.A. Nagarajan} and 
\author[Padova] {A. Vitturi}
\address[Catania]{ Dipartimento di Fisica and INFN, Catania, Italy}
\address[Legnaro]{Laboratori Nazionali di Legnaro, INFN, Italy }
\address[Padova]{ Dipartimento di Fisica and INFN, Padova, Italy}

\begin{abstract}
Quadrupole transition densities in neutron-rich nuclei in the vicinity
 of the neutron drip-line are calculated in the framework of the Random
 Phase Approximation.  The continuum is treated by 
 expansion in oscillator functions.
 We focus on the states which contribute to the usual Giant Quadrupole
 Resonance, and not on the low-lying strength which is also expected in 
 such nuclei and whose collective character is still under debate.
 We find that, due to the large neutron skin in these nuclei, the isoscalar
 and isovector modes are in general strongly mixed.  
 We further show that the transition densities corresponding to the GQR states
 can be reasonably well described by the collective model in terms of
 in phase and out of phase oscillations of neutron and proton 
 densities which have different radii.
\end{abstract}
\end{frontmatter}

During recent years, in view of the prospects of new dedicated radioactive
beam facilities, considerable interest has been focussed to the study of
the effect of neutron skin on the collective properties of neutron-rich
nuclei~\cite{1}. Largely the interest has been directed towards the possible
occurrence of sufficient multipole strength at low excitation energies in
such nuclei. It is now excepted that this feature is primarily a
consequence of the extremely weak binding of the neutrons close to the Fermi
surface, whereas the collective or single-particle nature of these
excitations is still unclear~\cite{2}.  It is also  of interest to study
neutron-rich nuclei which are still far from the neutron  drip-line, so
that the last nucleons are not so weakly bound, and ask what the effect of
the neutron skin is on the traditional collective oscillations of these
systems.  In view of the presence of the neutron skin one would expect an
enhanced mixing of the isoscalar and isovector modes.

\begin{figure}
\begin{center}
\includegraphics[bb= 110 110 550 719,angle=-90,scale=0.6]{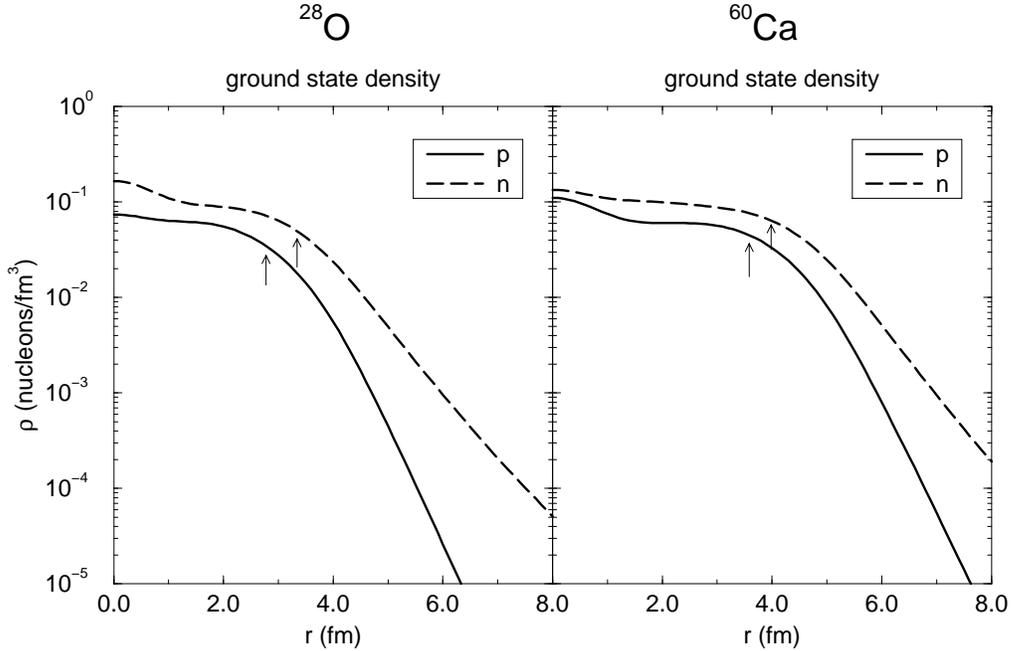}
\end{center}
\caption 
{ Hartree-Fock proton (solid line) and neutron (dashed line) 
ground-state densities for $^{28}$O and $^{60}$Ca. The interaction  used was
SGII.}     
\end{figure}

To investigate this point, we have performed microscopic calculations for
nuclei $^{28}$O and $^{60}$Ca, based on spherical Hartree-Fock model with
Skyrme SGII interaction.  These interaction predicts the last neutron to be
bound by 3.25 MeV in $^{28}$O and 5.1 MeV in $^{60}$Ca. The proton and
neutron densities for both $^{28}$O and $^{60}$Ca are shown in fig.~1. 
These nuclei display a neutron skin (the difference in the root-mean-square
radii amounts to about 0.6 fm and 0.4 fm, in the two cases, respectively),
without any significant neutron halo, since the last neutrons in these
nuclei are not too weakly bound.  As a further consequence of the
relatively large binding energy,  we do not expect any significant
multipole strength at very low excitation  energies in these nuclei 
\footnote { Different choices of the
Skyrme interaction may lead to lower binding energies of the last neutron
and may result in significant multipole strength at low   excitation  
energy~\cite{2,3} }.  In view of this, we believe that the continuum states  
can be adequately treated by expanding them in oscillator functions of
different principal quantum numbers,  a procedure which is cumbersome for a
system just on the drip line,  which would require an extremely large number
of oscillator functions. The adequacy of the oscillator expansion was
however tested by  varying the number of oscillator states and verifying
that the multipole strength distribution is not strongly affected for the
collective part of the response. The  strength distributions were
calculated for multipole operators of the form $\hat O_{\lambda
\mu}=r^{\lambda}Y_{\lambda \mu}(\hat r)$ for both neutrons and protons. 

The collective excitations of these nuclei were determined in the RPA, 
using the full residual interaction,
i.e. both the isoscalar and isovector components of
the Skyrme force were included simultaneously. The RPA states are therefore
in general connected to the ground state by both isoscalar and
isovector operators. 

\begin{figure}
\begin{center}
\includegraphics[bb= 80 80 550 719,angle=-90,scale=0.7]{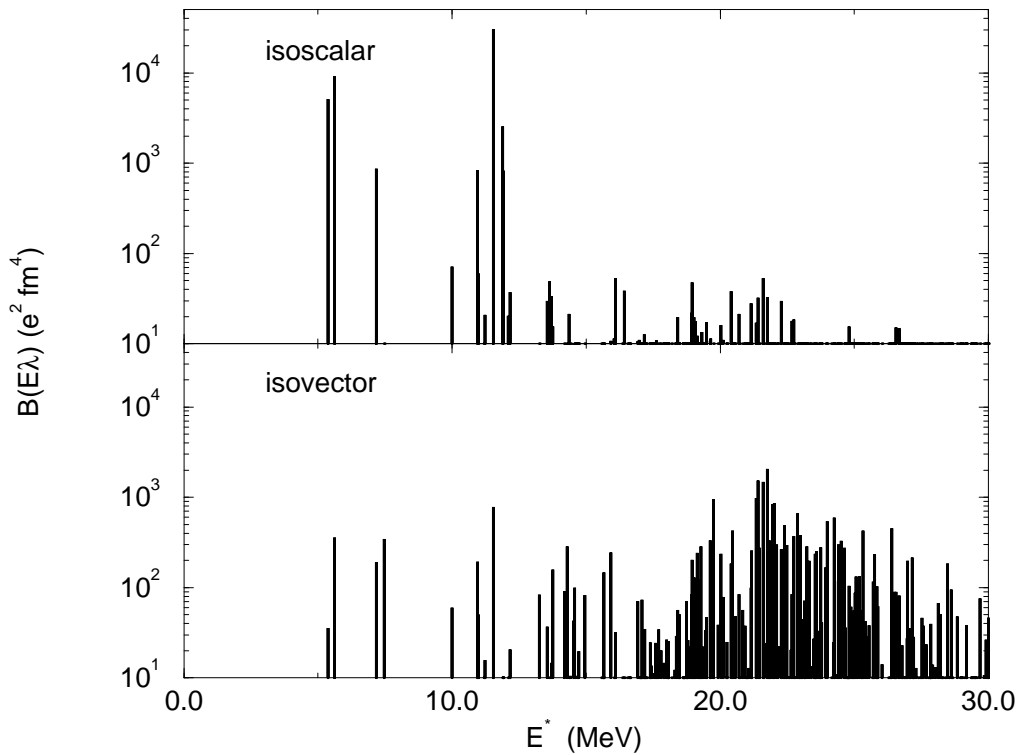}
\end{center}
\caption 
{Isoscalar and isovector strength distributions for the
RPA quadrupole states for the nucleus $^{208}$Pb. }     
\end{figure}

We already know that even for $\beta$-stable nuclei with $N \ge Z$  the
collective vibrations are mixtures of isoscalar and
isovector vibrations.  In the absence of neutron skin one expects the ratio
of  isovector and isoscalar amplitude for the isoscalar
giant resonance to be approximately equal to
$(N-Z)/A$.  As an example we show in fig.~2 the isoscalar and isovector
quadrupole response  for $^{208}$Pb. In correspondence with the
concentration of the isoscalar strength  around 11 MeV one also observes
some isovector strength of the order of 3~\% (namely close to ratio
$((N-Z)/A)^2$, which is 4~\%) while the main concentration of the isovector
strength occurs around 22 MeV.

\begin{figure}
\begin{center}
\includegraphics[bb= 100 100 550 719,angle=-90,scale=0.6]{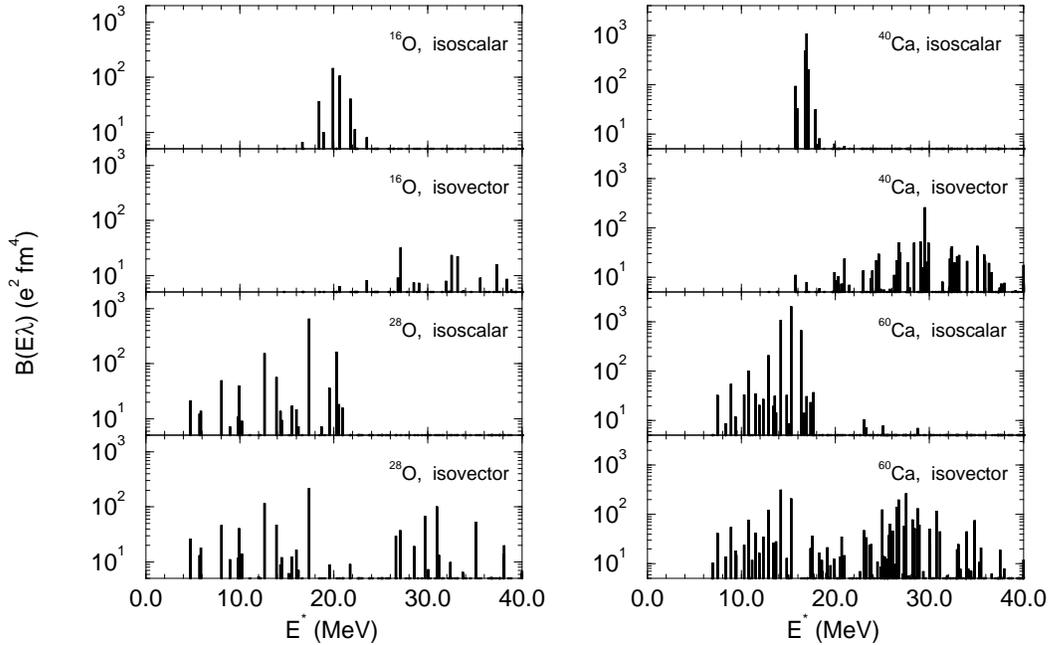}
\end{center}
\caption 
{Comparison of the isoscalar and isovector strength 
distributions for the RPA quadrupole states for $^{16}$O and $^{28}$O
(left panel) and for $^{40}$Ca and $^{60}$Ca (right panel).}     
\end{figure}

In view of the interest in the effect of the neutron excess on the isoscalar
and isovector response functions the RPA results for  $^{28}$O and
$^{60}$Ca are shown in fig.~3, compared with the analogous quantities for
$^{16}$O and $^{40}$Ca, respectively. In the case of the $N = Z$ nuclei
$^{16}$O and $^{40}$Ca the isoscalar and isovector distributions of
strength are rather well separated,  corresponding to pure isoscalar and
isovector excitations.  In contrast to this, one observes in the case
of $^{28}$O
and $^{60}$Ca firstly a large fragmentation of  the isoscalar strength  and
secondly a very large fragmentation of the isovector strength, down to the
region where there is a large isoscalar strength. In tables~1 and 2 the
$B(E2)$'s for certain representative RPA states for $^{28}$O and $^{60}$Ca
are shown, together with the percentage of the isoscalar and isovector sum
rules that each state exhausts. Note that the total EWSR for isoscalar and
isovector are different due to the exchange term in the latter~\cite{4}.
Note also that at variance with the situation for systems around the
$\beta$-stability, the proton and neutron EWSR's do not simply scale
according to $Z$ and $N$, due to the presence of the neutron skin which
leads to different rms radii for protons and neutrons.

\begin {table} 
\caption {  Properties of some selected RPA quadrupole states 
in $^{28}$O.  For each state we quote the excitation energy, fraction (in
percentage) of isoscalar and isovector EWSR exhausted by the state, $B(E2)$
values for the proton and neutron components, and corresponding isoscalar and
isovector values.}

\begin{tabular}{|r|r|r|r|r|r|r|} \hline

 Energy & EWSR &EWSR & $B(E2)_p$ & $B(E2)_n$ & $B(E2)_{is}$
 &$B(E2)_{iv}$ \\  
 (MeV) &  (\%)is & (\%)iv &($e^2 fm^4$) &($e^2 fm^4$) &($e^2 fm^4$) 
&($e^2 fm^4$)  \\   \hline\hline
 17.417 &  48.9 &  14.4  &  28.39 & 399.32 &  640.66 &  214.76 \\ \hline
 20.315 &  14.2 &   0.5  &  25.55 &  58.08 &  160.67 &    6.59 \\ \hline
 30.968 &   0.  &  12.7  &  24.18 &  30.47 &    0.36 &  108.94 \\ \hline

\end{tabular}

\end{table}

\begin {table} 
\caption {  As in Table~1, for the case of $^{60}$Ca. }

\begin{tabular}{|r|r|r|r|r|r|r|} \hline

 Energy & EWSR &EWSR & $B(E2)_p$ & $B(E2)_n$ & $B(E2)_{is}$
 &$B(E2)_{iv}$ \\  
 (MeV) &  (\%)is & (\%)iv &($e^2 fm^4$) &($e^2 fm^4$) &($e^2 fm^4$) 
&($e^2 fm^4$)  \\   \hline\hline
15.315 &  44.  &   4. &   241.11 & 912.03 & 2091.00 & 215.28 \\ \hline
16.373 &  15.5 &   0. &   154.88 & 188.45 &  685.01 &   1.64 \\ \hline
27.542 &   0.1 &   8. &    52.73 &  82.37 &    3.29 & 266.91 \\ \hline

\end{tabular}

\end{table}

The nature of the different states and the effect of the neutron skin are 
best evidenced by looking at the transition densities. To this end 
transition densities to the selected states in $^{28}$O and $^{60}$Ca listed 
in the tables~1 and 2 are
shown in Fig.~4. In each one of the frames, the isoscalar and
isovector transition densities along with the separate neutron and
proton transition densities are shown. 
The figure confirms that the isoscalar and isovector
modes are mixed and contribute to each state. 

\begin{figure}
\begin{center}
\includegraphics[bb= 10 70 550 719,angle=-90,scale=0.6]{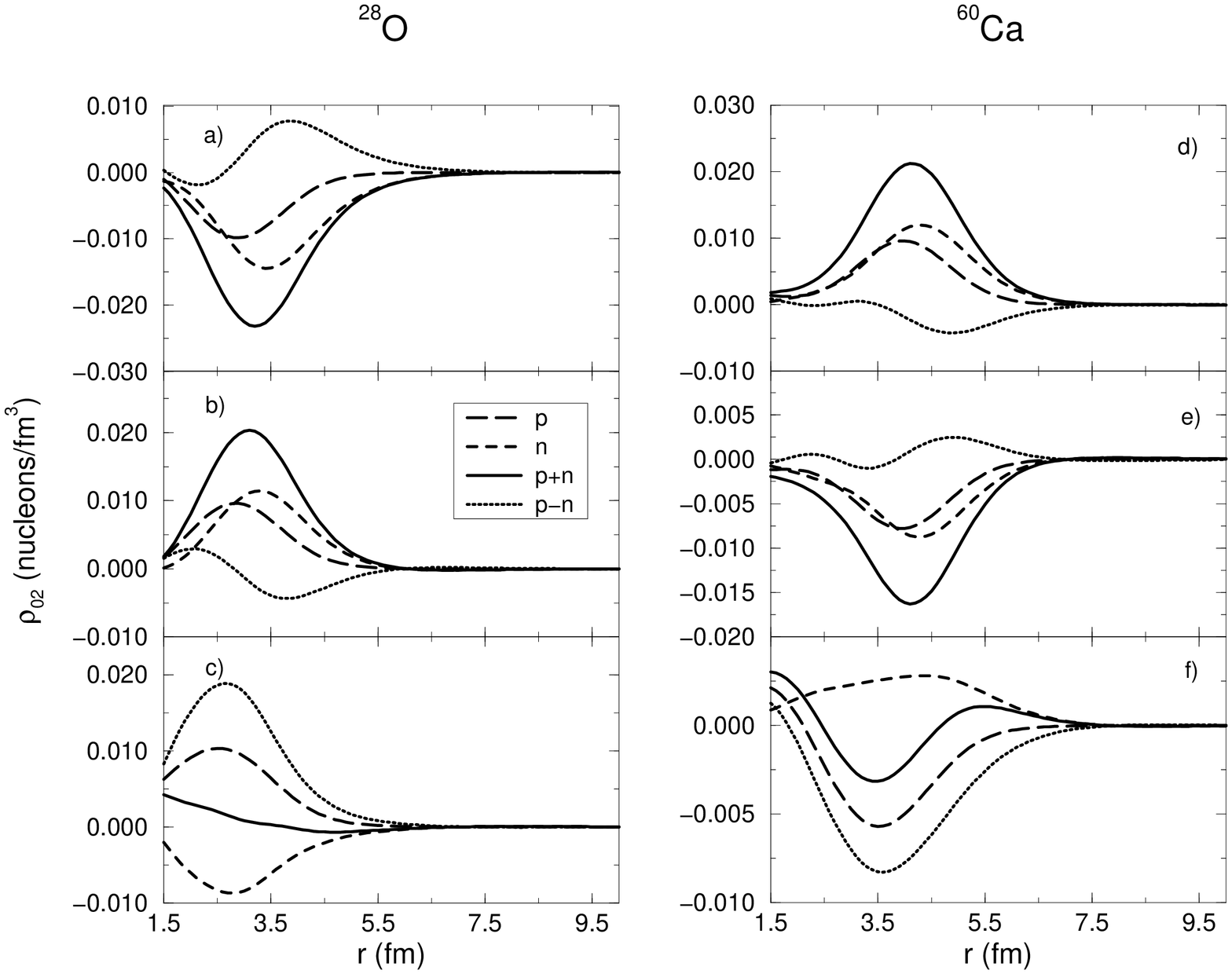}
\end{center}

\caption 
{RPA transition densities for the selected states 
in $^{28}$O listed in Table~1 are shown in the left panel.  The proton
and neutron transition densities are shown together with the isoscalar
and isovector transition densities.  The right panel shows the corresponding
quantities for the states listed in Table~2 for $^{60}$Ca.}     
\end{figure}

We can compare the corresponding transition densities with two
different prescriptions for the collective model.  In one of the 
prescriptions (Bohr and Mottelson model~\cite{5}), we assume that 
the change in the neutron and proton densities are proportional to 
their derivatives, according to
$$
\delta \rho _n~=\beta_2^n~R_n~{d \rho_n \over dr}
$$  
$$
\delta \rho _p~=\beta_2^p~R_p~{d \rho_p \over dr}
$$  
\noindent
where $\rho_n$ and $\rho_p$ are the Hartree-Fock  densities. 
In the second prescription we use the Tassie model~\cite{6}, 
where $\beta_2~R$  is replaced by $\beta_2~r$. 
By imposing that the collective transition densities lead to the same
$B(E2)$'s as the RPA ones and choosing $R_{n,p} = \sqrt{3 / 5} <r^2_{n,p}>$
we have obtained the $\beta$ values reported in Table~3. It is worthwhile
to notice that the $\beta$ values corresponding to the two collective model
prescriptions are very close to each other.

\begin {table} 
\caption {  Deformation parameters $\beta_2^n$ and $\beta_2^p$ obtained
within the collective Tassie and Bohr-Mottelson models, for each quadrupole 
state considered in Tables~1 and 2.  The values have been obtained from the
condition of yielding, within the collective models, the microscopic RPA
values, separately for neutrons and protons.}

\begin{tabular}{||l|r|r|r|r|r||} \hline
 Nuclei & Energy & \multicolumn{1}{c|}{$\beta_2^p$} &
 \multicolumn{1}{c|}  {$\beta_2^n$} & \multicolumn{1}{c|}
 {$\beta_2^p$} &  \multicolumn{1}{c||} {$\beta_2^n$} \\ 
  & (MeV)  &  Tassie   & Tassie    & B. \& M.  & B. \& M.  \\ \hline\hline 
  $^{28}O$
 &17.417 &   0.096  &   0.099  &   0.098  &   0.103 \\ \hline
 &20.315 &   0.091  &   0.038  &   0.093  &   0.039 \\ \hline
 &30.968 &   0.088  &  -0.027  &   0.091  &  -0.029 \\ \hline\hline
  $^{60}Ca$ 
 &15.315  &  0.068  &   0.053  &   0.069  &   0.054 \\ \hline
 &16.373  &  0.054  &   0.024  &   0.055  &   0.025 \\ \hline 
 &27.542  &  0.032  &  -0.016  &   0.032  &  -0.016 \\ \hline

\end{tabular}

\end{table}
 
\begin{figure}
\begin{center}
\includegraphics[bb= 80 80 600 719,angle=-90,scale=0.58]{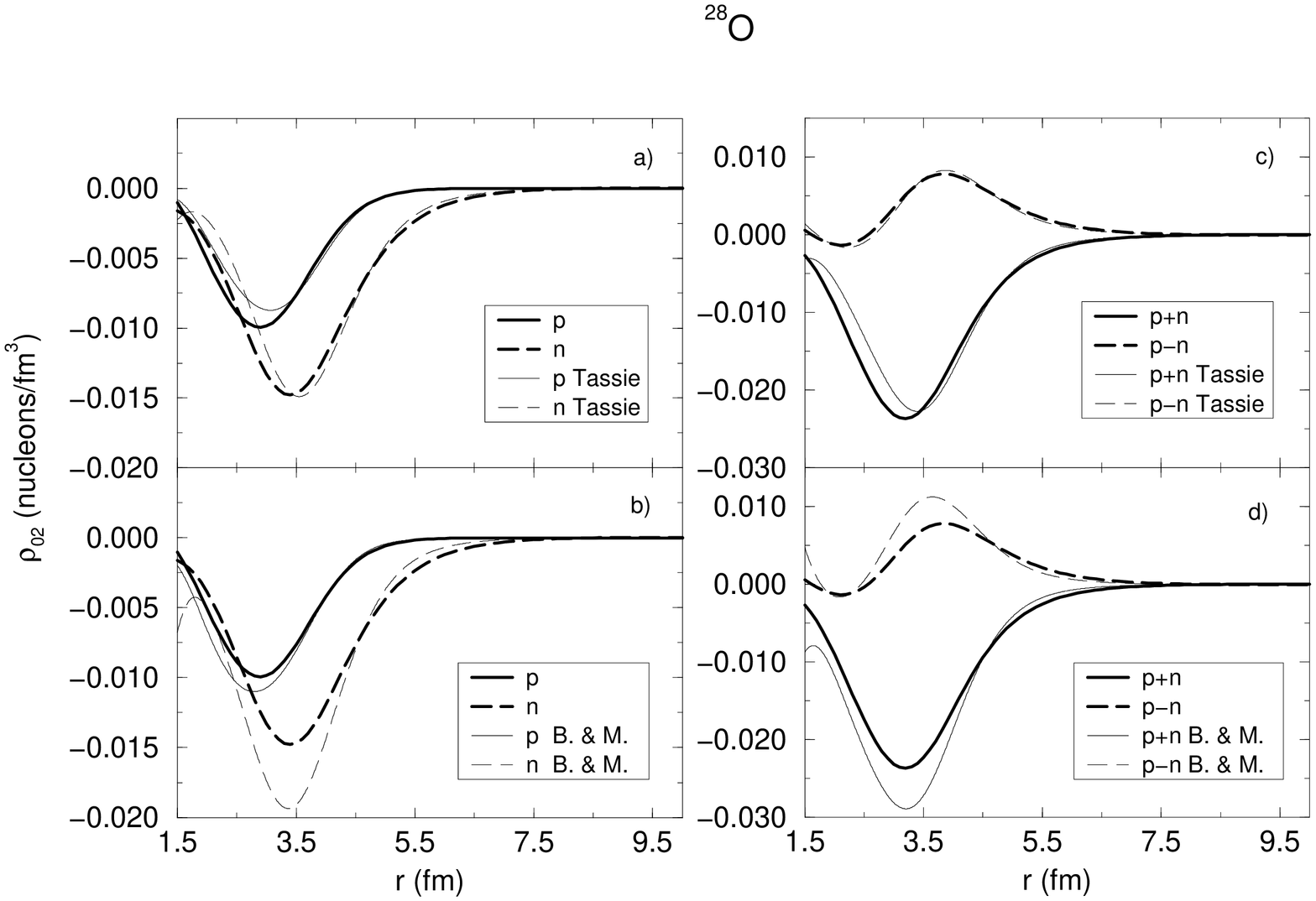}
\end{center}
\caption 
{Neutron and proton RPA transition densities for the 
quadrupole state at 17.4 MeV in $^{28}$O are compared with the
predictions of the Tassie model in Fig.~5a and with the predictions of
the  Bohr-Mottelson model in Fig.~5b. The corresponding isoscalar and
isovector transition densities are compared with the Tassie and
Bohr-Mottelson  models in Figs.~5c and 5d, respectively. }     
\end{figure}

As quoted in Table~1, the states at 17.4 MeV in $^{28}$O and at 15.3
MeV in  $^{60}$Ca exhaust approximately half of the isoscalar EWSR,
and  therefore can be associated with the usual collective Isoscalar
Giant  Quadrupole Resonance (GQR). This interpretation is confirmed by
the  fact that the $\beta$ values for neutrons and protons have the
same sign and practically the same magnitude, which in the collective
picture means that the proton and neutron densities $\rho_n$ and 
$\rho_p$ oscillate in phase and with the same amplitude. The
predictions of  the two models in the case of $^{28}$O for the neutron and
proton transition densities are shown in Fig.~5a,b along with the RPA results, 
while the corresponding isoscalar and isovector transition densities
are compared in Fig.~5c,d with the RPA results. Both collective model
predictions and the RPA transition densities seem to agree
qualitatively  with each other.
Similar agreement between the collective model predictions and the RPA 
transition densities was observed for the isoscalar collective state at 
15.3 MeV in $^{60}$Ca (see fig. 6).

\begin{figure}
\begin{center}
\includegraphics[bb= 80 80 600 719,angle=-90,scale=0.58]{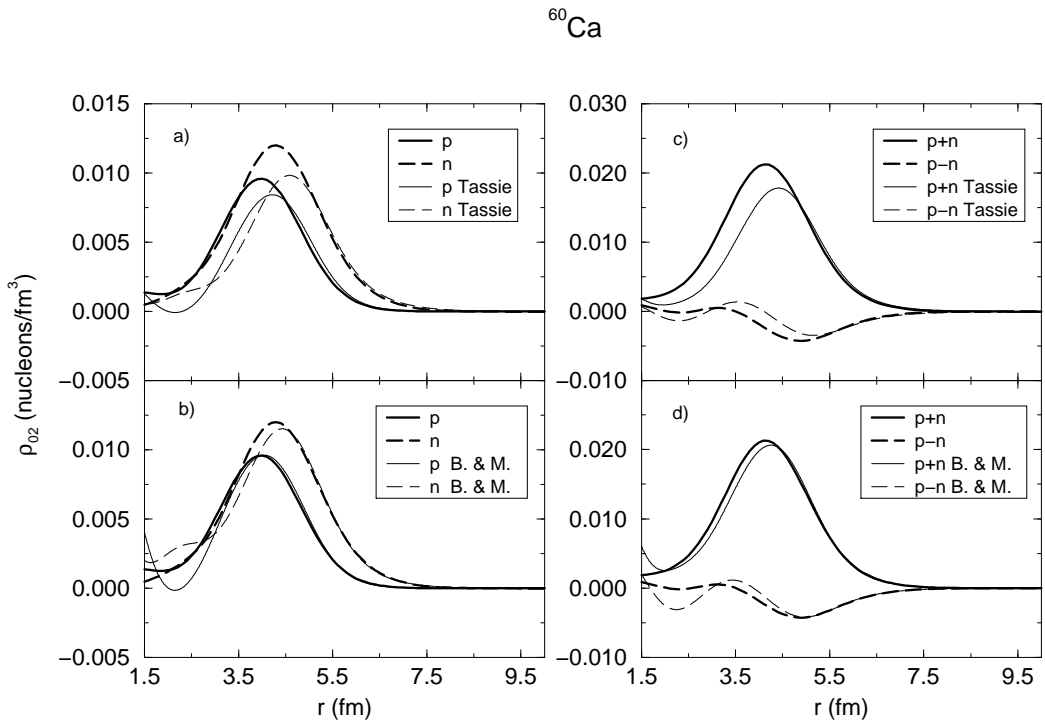}
\end{center}
\caption 
{ Same as in Fig.~5 for the state at 15.3 MeV in $^{60}$Ca.}     
\end{figure}

In fig.4c and 4f we have shown the transition densities for two
representative states which are dominantly of isovector
character. Again  in this case one can compare
the RPA results with the collective model.
In this case the $\beta$ values for neutrons and protons 
have opposite sign, corresponding to
the picture in which the neutron and proton
densities oscillate out of phase. 
As an example we show in Fig.~7 the comparison of the RPA transition
densities with those predicted by both Tassie and Bohr and Mottelson versions
of the collective model for the ``isovector'' state at 30.9 MeV in
$^{28}$O, with the value of $\beta$ reported in table 3 . 
Unlike the case of the isoscalar resonances (see Fig.~5), these isovector
states exhaust very small fraction of the EWSR.  It is not clear if their
description in terms of a collective model is appropriate.  It can be
seen  from the table that the corresponding $\beta$ parameters for the
neutrons and protons are quite different.  So they can only be
qualitatively interpreted in terms of the collective model.

\begin{figure}
\begin{center}
\includegraphics[bb= 80 80 600 719,angle=-90,scale=0.58]{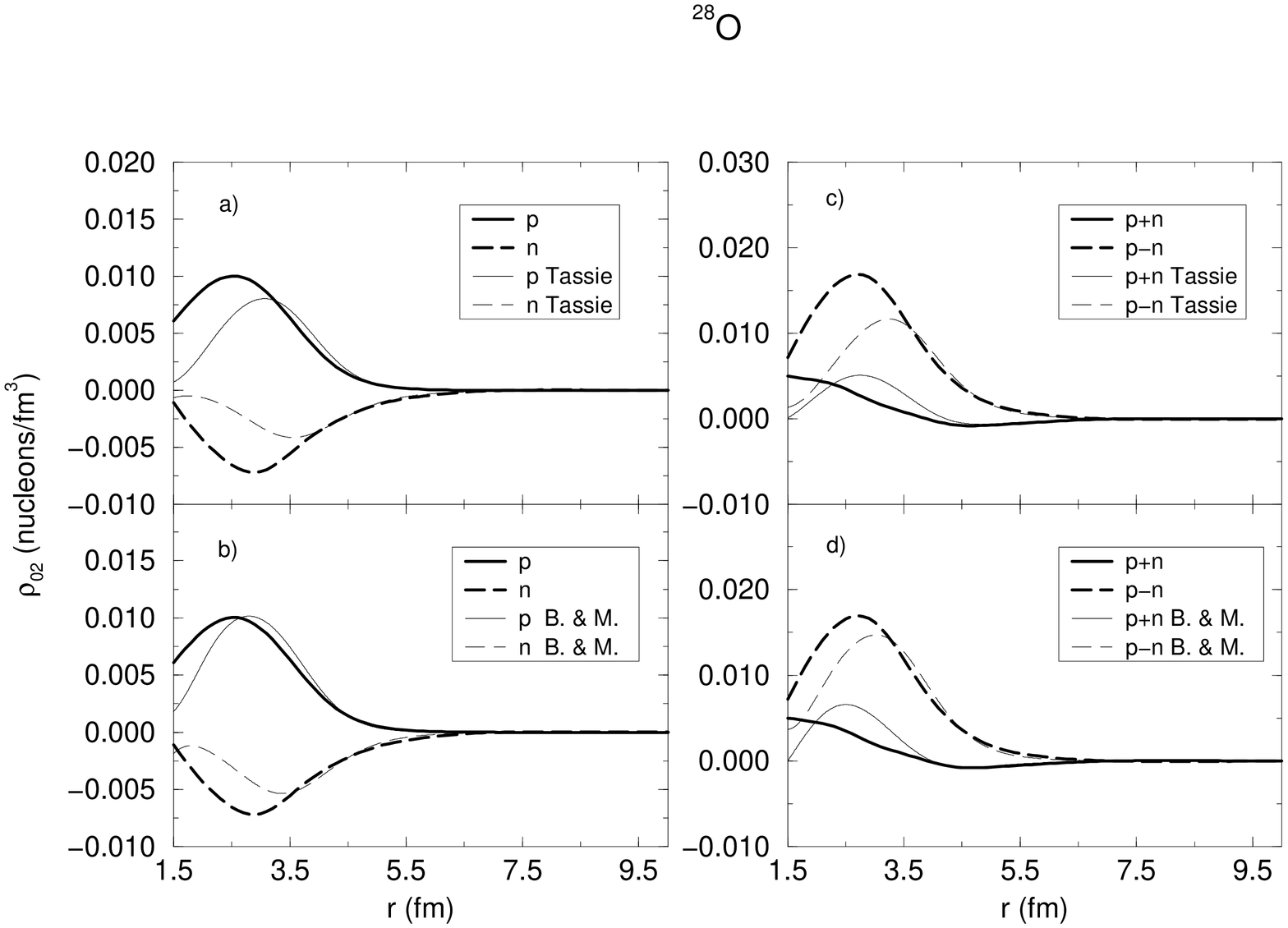}
\end{center}
\caption 
{  Same as in Fig.~5 for the dominantly isovector state 
at 30.9 MeV in $^{28}$O. }     
\end{figure}

The shape of the transition densities displayed in Fig.~4 can put in evidence
further aspects which are a direct consequence of the displacement of proton
and neutron radii and which do not necessarily show up in the integrated
$B(E2)$ values.  For the states of isoscalar character the proton and
neutron components to the transition densities peak in different positions,
thus leading to a surface-peaked  isoscalar transition density and
conversely to an isovector density which has a node on the 
surface \footnote  {Similar findings were reported by
Halbert and Satchler~\cite{7} in their analysis of proton inelastic 
scattering by $^{208}$Pb.}. As aconsequence, the isovector 
electromagnetic matrix elements are quenched  because of cancellation, 
and the isoscalar matrix elements dominate.  Yet, beyond the
nuclear surface, the neutrons give practically the only contribution, and
therefore isoscalar and isovector transition densities are of comparable
magnitude.  We therefore expect that these states, although of isoscalar
character, will respond in an equivalent way to isoscalar and
isovector heavy-ion probes that are only sensitive to the surface.
There may still be some effects of the nodes of the transition density
due to the finite range of the effective nucleon-nucleon interaction.
A similar statement may be valid for the isovector modes (cf. for
example Fig.~4f).  In this case it is the isoscalar transition density which 
displays a node on the surface and this leads to a small isoscalar matrix 
element.  Again, however, in the tail of the distribution, isoscalar and
isovector transition densities are comparable, and therefore we may expect
such a state to be also appreciably excited by a nuclear isoscalar field. 
One should, in general, expect to see the effect of the neutron skin in 
nuclear excitation which is sensitive to the behaviour of the transition 
densities near the nuclear surface.

To summarize, quadrupole transition densities as well as the quadrupole
response of neutron rich nuclei were investigated in the framework of
the RPA.  The transition densities clearly exhibit the effect of the
neutron skin in the case of neutron excess nuclei beyond the region of
$\beta$-stability.  The effect of the neutron skin is less apparent
in the case of the $B(E2)$'s since these involve an integral of the 
transition densities and the occurrence of nodes in the latter leads to
cancellation effects.  The results clearly indicate a strong mixing of
isoscalar and isovector strengths in the RPA states.  In the case of the nuclei
$^{28}$O and $^{60}$Ca, states were observed which exhaust around 40-50 \%
of the EWSR.  In these cases, the transition densities were seen to behave
like collective in-phase oscillations of the neutrons and protons with the
same amplitude.  The isovector strength, on the other hand, was observed to
be strongly fragmented with no single state carrying more than 10-12 \% of
the EWSR.  No realistic candidate for a collective isovector out-of-phase
oscillation of the densities was observed.

The effect of the neutron skin is expected to become apparent in nuclear
excitation where the details of the transition densities near the nuclear 
surface will be effective.  The detailed study of Coulomb and nuclear
inelastic excitation of the neutron-rich nuclei is therefore of interest 
and should be pursued in the future.

We acknowledge Ray Satchler for useful comments.


\begin{thebibliography}{99}

\bibitem{1} T. Otsuka, A. Muta, M. Yokoyama, N. Fukunishi and T. Suzuki,
          Nucl. Phys. A588 (1995) 113c  and references therein.
\bibitem{2} F. Catara, C.H. Dasso and A. Vitturi, 
          Nucl. Phys.  A602 (1996) 181;
          H. Sagawa, Nuyen van Giai, N. Yakigawa, M. Ishihara and K. Yazaki,
          Z. Phys. A351  (1995) 385.
\bibitem{3} I. Hamamoto and H. Sagawa, Phys. Rev. C53 (1996) R1492; 
          F. Ghielmetti, G. Col\`o, E. Vigezzi, P.F. Bortignon and
          R.A. Broglia, Phys. Rev.  C (to appear in November 1996 issue).
\bibitem{4} E. Lipparini and S. Stringari, Phys. Reports 175C  (1989) 103.
\bibitem{5} A. Bohr and B.R. Mottelson, {\em Nuclear Structure}, Vol.~2,
          Benjamin (New York) 1975.
\bibitem{6} L.J. Tassie. Austr. J. Phys. 9 (1956) 407. 
\bibitem{7} E.C. Halbert and G.R. Satchler, Nucl. Phys. A233 (1974) 265 . 

\end{thebibliography}
\end{document}